\documentclass[twocolumn,prb,amsmath,amssymb,showpacs,superscriptaddress]{revtex4}
\usepackage{graphicx}

\begin{document}

\title{Tunnel-barrier-enhanced dc voltage signals induced by magnetization dynamics in magnetic tunnnel junctions}

\author{Yaroslav Tserkovnyak}
\affiliation{Department of Physics and Astronomy, University of California, Los Angeles, California 90095, USA}
\author{T. Moriyama}
\author{John Q. Xiao}
\affiliation{Department of Physics and Astronomy, University of Delaware, Newark, Delaware 19716, USA}

\date{\today}

\begin{abstract}
We theoretically study the recently observed tunnel-barrier-enhanced dc voltage signals generated by magnetization precession in magnetic tunnel junctions. While the spin pumping is suppressed by the high tunneling impedance, two complimentary processes are predicted to result in a sizable voltage generation in ferromagnet (F)$\mid$insulator (I)$\mid$normal-metal (N) and F$\mid$I$\mid$F junctions, with one ferromagnet being resonantly excited. Magnetic dynamics in F$\mid$I$\mid$F systems induces a robust charge pumping, translating into voltage in open circuits. In addition, dynamics in a single ferromagnetic layer develops longitudinal spin accumulation inside the ferromagnet. A tunnel barrier then acts as a nonintrusive probe that converts the spin accumulation into a measurable voltage. Neither of the proposed mechanisms suffers from spin relaxation, which is typically fast on the scale of the exponentially slow tunneling rates. The longitudinal spin-accumulation buildup, however, is very sensitive to the phenomenological ingredients of the spin-relaxation picture.
\end{abstract}


%72.15.Gd Galvanomagnetic and other magnetotransport effects
%76.50.+g Ferromagnetic, antiferromagnetic, and ferrimagnetic resonances; spin-wave resonance
%72.25.Mk Spin transport through interfaces
%72.25.Ba Spin polarized transport in metals

\maketitle

Voltage induced by magnetization dynamics in layered ferromagnet$\mid$normal metal (F$\mid$N) structures has recently attracted considerable attention as one of the basic building blocks in magnetoelectronics.\cite{bergerPRB99,costachePRL06,wangPRL06vg,chuiPRB08,xiaoPRB08} Much of the interest in the problem was motivated by the magnetically ``pumped" spin flows,\cite{tserkovPRL02sp,tserkovPRB02sp,tserkovRMP05} which are expected to generate detectable voltage signals in magnetic multilayers.\cite{brataasPRB02} The latter can provide a direct manifestation of the spin-pumping effect, as well as a potentially useful electric probe for magnetic dynamics. A recent measurement\cite{costachePRL06} of the voltage signal of the order of 100~nV in an Ohmic permalloy (Py)$\mid$aluminum (Al) structure appears to be well explained by the spin-pumping mechanism.\cite{wangPRL06vg} In contrast, a more recent experiment\cite{moriyamaPRL08} on a Py$\mid$Al-based structure with an Al$_2$O$_3$ tunnel-barrier interlayer between Py and Al films reported a surprisingly large signal (of the order of 1~$\mu$V) at a smaller resonance frequency. This appears to suggest a different mechanism for voltage generation, since, if anything, the tunnel barrier is expected to suppress the spin pumping and the ensuing voltage. It is useful to recall that a sizable spin pumping requires good interfacial transparency, while the induced voltage is established by an interplay between the pumped spin-injection and spin-relaxation rates.\cite{brataasPRB02} Even a thin tunnel barrier is in practice sufficiently opaque to strongly suppress the spin-pumping induced voltage.

In this Communication, we investigate alternative scenarios for the tunnel-barrier-enhanced voltage signals in F$\mid$I$\mid$N and F$\mid$I$\mid$F systems. We propose two mechanisms for voltage  generation|with one being effective in F$\mid$I$\mid$N structures, while both interplay on equal footing in F$\mid$I$\mid$F junctions. In spite of certain qualitative differences, our picture is conceptually reminiscent of the spin-pumping physics.\cite{tserkovPRL02sp} Parts of our theory concerning F$\mid$I$\mid$F junctions are closely related to Ref.~\onlinecite{xiaoPRB08}, but based on a very different and more phenomenological approach. Our mechanism for voltage generation in F$\mid$I$\mid$N structures does not appear to have much in common with that developed in Ref.~\onlinecite{chuiPRB08}, which is based on the tunneling spin pumping and the interplay between spin diffusion and self-consistent screening near the junction. The central stage in our theory will be given to direct charge pumping, rather than spin pumping that gets subsequently converted into a voltage signal.\cite{brataasPRB02,chuiPRB08}

\begin{figure}
\centerline{\includegraphics[width=0.8\linewidth]{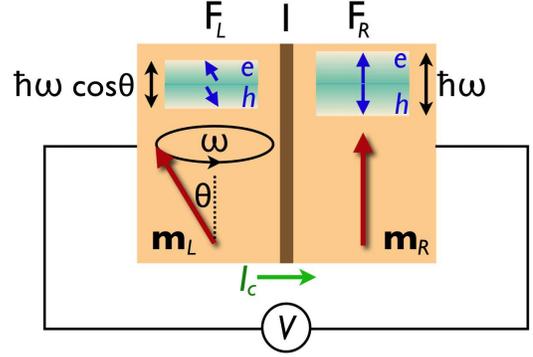}}
\caption{(Color online) Voltage generated in an F$_L$$\mid$I$\mid$F$_R$ junction by magnetically induced charge pumping $I_c$. In the absence of spin relaxation, magnetic precession builds up a spin imbalance of $\hbar\omega$ in F$_R$ and $\hbar\omega\cos\theta$ in F$_L$, along the respective magnetization directions. As shown in the text, this must necessarily be accompanied by a charge pumping for a finite $\theta$. In a realistic situation where the spin-relaxation rate is much faster than the tunneling injection rate, we can safely neglect the spin-pumping component and calculate the resulting voltage signal $V$ induced by the charge pumping $I_c$ alone.}
\label{fig1}
\end{figure}

We start by considering the F$_L$$\mid$I$\mid$F$_R$ spin valve sketched in Fig.~\ref{fig1}. To bring out the key physics, it is sufficient to treat the following simple model Hamiltonian for itinerant electrons coupled to collective magnetic dynamics:
\begin{equation}
\hat{H}(t)=p^{2}/2m+V(\mathbf{r})+(\Delta/2)\mathbf{m}(\mathbf{r},t)\cdot\hat{\boldsymbol{\sigma}}\,,
\label{H}
\end{equation}
where $\mathbf{m}(\mathbf{r},t)$ is a unit vector pointing along the local magnetization direction, $\hat{\boldsymbol{\sigma}}$ is the vector of Pauli matrices, and the potential $V(\mathbf{r})$ includes crystal, disorder, and a possible external electric-field potential. $\Delta$ is the material-dependent exchange field, which for simplicity is taken to be the same in both ferromagnets. We will take $\mathbf{m}_L(t)$ in the left ferromagnet to be spatially uniform and steadily precessing around the $z$ axis, with a constant cone angle $\theta$. The right ferromagnet is stationary, pointing along the $z$ axis: $\mathbf{m}_R\equiv\mathbf{z}$. Otherwise, the magnetic tunnel junction will be treated as mirror symmetric.

A steady precession of $\mathbf{m}_L$ around $\mathbf{m}_R$ modulates spin-dependent tunneling matrix elements, which may allow electron pumping across the barrier in the absence of any external bias. In the adiabatic limit assumed in the following, the pumping strength is proportional to the frequency of the magnetic precession. As electrons carry both spin and charge, the pumping can in general consist of spin and charge components. In the case of an F$\mid$I$\mid$N junction, the pumping into the normal metal turns out to be of pure spin character with a vanishing charge component.\cite{tserkovPRL02sp} In contrast, we will see that the pumping current across the magnetic F$_L$$\mid$I$\mid$F$_R$ junction has a nonvanishing spin and charge admixture. Since in realistic metallic junctions the spin-relaxation rate usually overwhelms the tunneling injection rate, we can safely neglect the spin pumping and retain only the charge current component of the pumping process. In this regard, our picture is qualitatively different from Refs.~\onlinecite{brataasPRB02,wangPRL06vg}, which focused on how spin pumping is converted into a charge signal by ferromagnetic spin filtering in Ohmic multilayers. In the following, we will disregard spin pumping altogether, and we will devote our full attention to the direct charge pumping instead.

The simplest way to compute the dynamically induced pumping currents is to apply a spin-rotation transformation to Hamiltonian (\ref{H}) and solve for the equilibrium state in the reference frame, where the F$_L$ magnetization is static.\cite{tserkovPRB05} A snapshot of such steady-state solution in the laboratory frame of reference with Hamiltonian (\ref{H}) will in general look instantaneously out of equilibrium as manifested by some spin and charge buildups. In a steady state, the ensuing tunneling backflow currents should exactly cancel the initial pumping.\cite{tserkovRMP05,tserkovPRB05} One subtle but crucial point needs to be clarified here. The adiabatic pumping flows develop over the ferromagnetic coherence length of $\hbar v_F/\Delta$, which is atomistically short in transition-metal ferromagnets. As long as the spin-relaxation length in the ferromagnets is much longer, it will have no consequences for the strength of the pumped  spin and charge flows. We can thus proceed, for convenience, to compute the pumping currents in the absence of any spin relaxation. The hypothetical steady state thus acquires a finite spin buildup, as sketched in Fig.~\ref{fig1}, although in practice, any spin accumulation will be strongly reduced by spin relaxation.\cite{brataasPRB02} The following considerations, which are intricately based on manipulating finite spin buildups, should thus be viewed as only a ``trick" to calculate pumping flows with no implications for the spin accumulation that will eventually be reached. The underlying reasoning is that as long as the dynamics are sufficiently slow (on the scale of $\Delta^{-1}$), the strength of the pumped spin and charge flows is not related to the actual steady-state solution, which is established by balancing these flows with other transport and relaxation processes, or, in fact, to whether the steady state will be achieved at all.

For $\mathbf{m}_L$ precessing clockwise around $\mathbf{m}_R$, the transformed time-independent Hamiltonian is given by
\begin{equation}
\hat{H}^{\prime}=\hat{R}^{\dagger}\hat{H}\hat{R}-i\hbar\hat{R}^{\dagger}\partial_{t}\hat{R}=\hat{H}(0)-(\hbar\omega/2)\hat{\sigma}_{z}\,,
\label{Hp}
\end{equation}
where $\hat{R}=e^{-i\omega t\hat{\sigma}_z/2}$ is the spin-rotation transformation around the $z$ axis. The last term in Eq.~(\ref{Hp}) corresponding to the generator of the transformation $\hat{R}$ induces a spin splitting of $\hbar\omega$ in the right ferromagnet and a smaller splitting of $\hbar\omega\cos\theta$ in the left, dynamic ferromagnet, along the direction of its exchange field. It is this difference between spin splittings that drives the spin and charge pumping across the barrier. Note that the transverse field of $\hbar\omega\sin\theta$ in F$_L$ will affect the pumping only at the order of $\omega^2$, which is neglected in our adiabatic description. The adiabatic approximation is adequate as long as the frequency $\omega$ is much smaller than the exchange $\Delta$, as well as the rates of all the relevant processes, including spin relaxation. The charge current $I_{R\to L}$ due to the nonequilibrium spin accumulation of $\hbar\omega$ in F$_R$ is given by
\begin{align}
I_{R\to L}=&e|T|^2\frac{\hbar\omega}{2}D_\uparrow\left(D_\uparrow\cos^2\frac{\theta}{2}+D_\downarrow\sin^2\frac{\theta}{2}\right)\nonumber\\
&-e|T|^2\frac{\hbar\omega}{2}D_\downarrow\left(D_\uparrow\sin^2\frac{\theta}{2}+D_\downarrow\cos^2\frac{\theta}{2}\right)\nonumber\\
=&e|T|^2\frac{\hbar\omega}{2}\left(D_\uparrow^2-D_\downarrow^2\right)\cos^2\frac{\theta}{2}\,,
\label{Ib}
\end{align}
where $e$ is the negative electron charge, $T$ is the orbital part of the tunneling matrix element (assumed for simplicity to be spin independent), and $D_s$ is the spin-$s$ density of states, which is taken to be the same in two ferromagnets. Since the spin buildup is smaller by $\cos\theta$ in F$_L$, we find for the net charge pumping across the junction
\begin{equation}
I_c=(1-\cos\theta)I_{R\to L}=e|T|^2\frac{\hbar\omega}{2}\left(D_\uparrow^2-D_\downarrow^2\right)\frac{\sin^2\theta}{2}\,.
\label{Ic}
\end{equation}

The ``spin bias" of $\hbar\omega(1-\cos\theta)$ across the tunnel barrier, which drives this current, can also be physically interpreted as the difference in the Berry phase accumulation rates of the two spin directions adiabatically following the dynamic magnetization $\mathbf{m}_L(t)$. To this end, recall\cite{berryPRSLA84} that the solid angle $\Omega=2\pi(1-\cos\theta)$ enclosed by $\mathbf{m}_L$ upon a cycle of precession corresponds to the Berry phase $\pm\Omega/2$ for spins up (down). Magnetic precession in the left ferromagnet thus leads to the effective potential $\varphi_s=s\hbar\dot{\Omega}/2=s\hbar\omega(1-\cos\theta)/2$ for spins $s=\uparrow,\downarrow$. The resulting spin bias $\varphi_\uparrow-\varphi_\downarrow=\hbar\omega(1-\cos\theta)$ in turn drives the charge pumping (\ref{Ic}), providing an alternative physical picture for the effect.

The last step in our treatment of magnetic F$_L$$\mid$I$\mid$F$_R$ tunnel junctions is to divide the pumped charge current (\ref{Ic}) by the angle-dependent junction conductance $G(\theta)$, in order to compute the voltage $V=I_c/G$. Using
\begin{equation}
G=e^2|T|^2\left[\left(D_\uparrow^2+D_\downarrow^2\right)\cos^2\frac{\theta}{2}+2D_\uparrow D_\downarrow\sin^2\frac{\theta}{2}\right]\,,
\end{equation}
we find a simple relation for the voltage:
\begin{equation}
V=\frac{\hbar \omega}{2e} \frac{P\sin^2\theta}{1+P^{2}\cos\theta}\,,
\label{V}
\end{equation}
where
\begin{equation}
 P =(D_\uparrow-D_\downarrow)/(D_\uparrow+D_\downarrow)
\end{equation}
is the ferromagnetic polarization. The voltage (\ref{V}) vanishes in the trivial limits: $P=0$ (no magnetism) or $\theta=0$ (no precession). Note that the voltage is larger near the antiparallel alignment ($\theta\to\pi$) than near the parallel alignment ($\theta\to0$) by the factor of $(1+P^{2})/(1-P^{2})$, for the same precessional cone angle. The voltage reaches its maximum possible value of $\hbar\omega$ when $P=1$ and $\theta\to\pi$. This could lead to significant signals at modest precession angles near the antiparallel alignment if one utilizes a tunnel barrier with $P\approx1$ such as MgO. We believe that the above discussion is consistent with the approach developed in Ref.~\onlinecite{xiaoPRB08}, despite a different treatment of the insulating barrier: We assume tunneling Hamiltonian, while Ref.~\onlinecite{xiaoPRB08} considers specular interface scattering.

Similar considerations for the F$\mid$I$\mid$N tunneling give
\begin{align}
I_{R\to L}=&e|T|^2\frac{\hbar\omega}{2}D_N\left(D_\uparrow\cos^2\frac{\theta}{2}+D_\downarrow\sin^2\frac{\theta}{2}\right)\nonumber\\
&-e|T|^2\frac{\hbar\omega}{2}D_N\left(D_\uparrow\sin^2\frac{\theta}{2}+D_\downarrow\cos^2\frac{\theta}{2}\right)\nonumber\\
=&e|T|^2\frac{\hbar\omega}{2}D_N\left(D_\uparrow-D_\downarrow\right)\cos\theta\,,
\end{align}
where $D_N$ is the normal-metal density of states per spin. This is exactly equal to $I_{L\to R}$ due to the spin accumulation of $\hbar\omega\cos\theta$ in the F layer, so that $I_c=I_{R\to L}-I_{L\to R}=0$, as expected in general for F$\mid$N junctions.\cite{tserkovPRL02sp} Magnetic dynamics in an F$_L$$\mid$I$\mid$N$\mid$F$_R$ system would thus not induce any charge current and the associated voltage, as long as the quantum-size effects\cite{xiaoPRB08} are disregarded. The spin-current pumping through tunnel barriers, furthermore, should not produce a significant voltage signal, since the associated spin buildups decay fast on the scale of the exponentially low injection rate.\cite{brataasPRB02} In Ref.~\onlinecite{moriyamaPRL08}, however, the measured voltage across an F$\mid$I$\mid$N junction seemed to be enhanced rather than suppressed by the tunnel barrier. In the following, we argue that this voltage is likely to be induced not by the spin pumping across the tunnel barrier but rather by a nonequilibrium spin buildup intrinsic to the excited ferromagnet. The tunnel barrier then simply acts as a nonintrusive probe, transforming the spin accumulation into voltage.

\begin{figure}
\centerline{\includegraphics[width=0.7\linewidth]{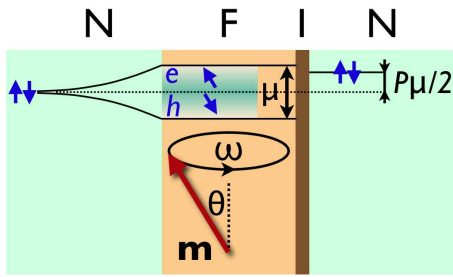}}
\caption{(Color online) Voltage measured in an F$\mid$I$\mid$N junction. In this case, charge pumping across the tunnel barrier vanishes, rendering the mechanism depicted in Fig.~\ref{fig1} ineffective. Depending on the model of spin relaxation in the ferromagnet, however, a tunnel barrier may detect a voltage signal due to the nonequilibrium spin buildup inside the precessing magnet. According to Eqs.~(\ref{ds0}) and (\ref{dsl}), the corresponding spin accumulation may, under the most favorable circumstances, be at most $\mu\sim\hbar\omega\cos\theta$, in the case of magnetic impurities adiabatically following the magnetization precession. This spin accumulation is measurable\cite{jedemaNAT01} by the tunnel barrier with the conductance polarization of $P$ as a voltage $P\mu/2$. The normal metal on the left mimics an Ohmic contact, which sets the reference potential at the spin-averaged electrochemical potential of the ferromagnet. Also schematically plotted on the left is the spin-diffusion profile of the spin-dependent electrochemical potential.}
\label{fig2}
\end{figure}

We proceed by considering an isolated ferromagnetic layer with a homogeneous and steadily precessing magnetization vector. In an idealized case with no spin relaxation, the rotating-frame arguments would predict that the magnetic dynamics induce a fictitious spin splitting of $\hbar\omega$ along the axis of precession. In practice, however, the actual spin accumulation turns out to be extremely sensitive to spin-relaxation processes in the ferromagnet, as detailed in the following. Let us demonstrate this with the use of the phenomenological Bloch equation
\begin{equation}
\frac{d\mathbf{s}}{dt}=\frac{\Delta}{\hbar}\mathbf{m}\times\mathbf{s}-\frac{(\mathbf{s}\cdot\mathbf{m})\mathbf{m}+s_0\mathbf{m}}{T_1}-\frac{\mathbf{m}\times\mathbf{s}\times\mathbf{m}}{T_2}\,,
\label{B}
\end{equation}
for the itinerant electron spin density $\mathbf{s}$. The first term on the right-hand side of Eq.~(\ref{B}) describes spin precession in the exchange field according to Hamiltonian (\ref{H}), the other two terms are the longitudinal and transverse spin relaxations, respectively. $-s_0\mathbf{m}$ is the equilibrium spin density corresponding to the static exchange splitting $\Delta$ along $\mathbf{m}$. Solving this equation in the rotating frame of reference yields (after a bit of straightforward algebra)
\begin{equation}
\frac{\delta s}{s_0}\approx\left(\frac{\hbar\omega}{\Delta}\right)^2\frac{T_1}{T_2}\sin^2\theta\,,
\label{ds0}
\end{equation}
where we assumed that $\omega,T_1^{-1},T_2^{-1}\ll\Delta/\hbar$, which is usually the case in practice. The longitudinal spin density $\delta s$ corresponds to the spin accumulation $\mu\approx(\delta s/s_0)\Delta\sim(\hbar\omega\sin\theta)^2/\Delta$ along $\mathbf{m}$, which is readily detectable by a voltage probe through a tunnel junction. \cite{jedemaNAT01} Figure~\ref{fig2} shows schematically how the tunnel barrier converts this spin accumulation into a measurable voltage $V\approx P\mu/2e$ in our model. This signal, however, is going to be very small since it scales quadratically with $\hbar\omega/\Delta\lesssim10^{-3}$ (at typical microwave frequencies and transition-metal exchange fields).

Note that Eq.~(\ref{B}) implicitly assumes that the disorder (causing spin relaxation) is quenched in the laboratory frame of reference, which is reflected in the form of the Bloch decay of the nonequilibrium spin density toward the instantaneous equilibrium value. However, we can also envision a scenario where some fraction of a hypothetical magnetic disorder (stemming, e.g., from lattice imperfections) is driven by the magnetic stiffness in response to collective ferromagnetic precession, so that it is partially ``quenched" in the rotating frame of reference. In a crude model where a fraction $\eta\leq1$ of the disorder follows the magnetic dynamics, while the remaining fraction $1-\eta$ is effectively stationary, we find for the spin accumulation
\begin{equation}
\mu\sim\eta\hbar\omega\cos\theta\,,
\label{dsl}
\end{equation}
along the instantaneous magnetization direction $\mathbf{m}$. We derived this result by transforming Eq.~(\ref{B}) into the rotating frame of reference, replacing $d\mathbf{s}/dt\to d\mathbf{s}/dt+\omega\mathbf{z}\times\mathbf{s}$ on the left-hand side. In addition, due to the disorder fraction $\eta$, which is taken to be stationary in the rotating frame of reference, we added the appropriate relaxation toward the spin density $(\mathbf{z}\,\hbar\omega/\Delta-\mathbf{m})s_0$. The extreme limit of $\mu=\hbar\omega\cos\theta$ when $\eta=1$ thus simply reflects the $\hbar\omega$ spin splitting along the $z$ axis in the rotating frame of reference. We may generally expect that the fudge parameter $\eta$ is angle dependent with $\eta\to0$ as $\theta\to0$ (since small-angle precession should have no appreciable effect on the disorder configuration). Within this picture, the measured voltage\cite{moriyamaPRL08} as a function of $\theta$ essentially traces out the function $\eta(\theta)$, which is a property of the dynamic response of the ferromagnetic spin impurities. The measured voltage\cite{moriyamaPRL08} certainly exceeds the spin accumulation (\ref{ds0}) by several orders of magnitude, while it has linear scaling with frequency and has an order of magnitude consistent with Eq.~(\ref{dsl}). The spin-relaxation properties of out-of-equilibrium ferromagnets, however, need to be better understood before making a concrete connection between the magnetic dynamics and the generated spin accumulation and voltage.

Tunnel barriers inserted in magnetic multilayers thus facilitate voltage generation in at least two different ways. On the one hand, F$\mid$I$\mid$F tunnel barriers support charge pumping inducing a detectable voltage, while on the other hand, tunnel contacts efficiently convert into voltage the nonequilibrium spin accumulations generated in the ferromagnets by their internal dynamics. The latter process can contribute to the voltage signals produced by F$\mid$I$\mid$N junctions, as those measured in Ref.~\onlinecite{moriyamaPRL08}, while both processes likely interplay in developing voltage signals across F$\mid$I$\mid$F junctions. It is important to note one qualitative difference between these two contributions to voltage generation: The charge-pumping voltage (\ref{V}) changes its sign if we flip the direction of either of the ferromagnets (since then, in our convention, the precession will change from clockwise to counterclockwise), while the voltage corresponding to the spin accumulation sketched in Fig.~\ref{fig2} should clearly be symmetric under the magnetic reversal.

Macroscopic magnetic moments can produce a variety of interesting dynamic phenomena. Resonantly exciting the magnetic moment in a ferromagnetic resonance, which can induce parametric spin pumping, is an example of one effect which has generated a great deal of interest over the past few years.\cite{tserkovRMP05} However, due to spin relaxation, spin currents produced by  these rotating magnetic moments are mostly localized to within the regions close to ferromagnetic interfaces. Opaque interfaces, such as in the case of tunnel barriers, would furthermore suppress spin-pumping effects. In this Communication, we showed that two ferromagnets in direct contact can dynamically induce also a charge pumping, which in turn generates robust voltage signals even in magnetic tunnel junctions in the presence of a fast spin relaxation. Any normal-metal interlayers, however, would strongly suppress the charge pumping and the associated voltage. In addition, we argued that when magnetic disorder is modulated by the ferromagnetic dynamics, the latter is accompanied by a nonequilibrium spin accumulation, which produces voltage drops across the adjacent tunnel barriers.

We acknowledge numerous helpful discussions with B.~K. Nikoli{\'c}. This work was supported in part by NSF DMR Grant 
No.~0405136 and DOE Grant. No.~DE-FG02-07ER46374.

\end{document}